\documentclass[aps, prd, superscriptaddress, nofootinbib, preprintnumbers]{revtex4-2}
\usepackage[colorlinks=true]{hyperref}
\usepackage{xcolor}
\hypersetup{colorlinks=true, citecolor=blue, urlcolor=blue, linkcolor=blue}
\usepackage{amsmath,amssymb}
\usepackage[final]{graphicx}
\usepackage{subcaption}
\usepackage[belowskip=0.5pt,aboveskip=0.5pt]{caption}
\usepackage{float}
\usepackage{adjustbox}
\usepackage {ulem}

\begin{document}

\title{The electromagnetic form factors in the $N_{f}=4$ holographic QCD}

\author{Hiwa A. Ahmed}
\email{hiwa.a.ahmed@mails.ucas.ac.cn}
\affiliation{School of Nuclear Science and Technology, University of Chinese Academy of Sciences, Beijing, 100049, P.R. China}
\affiliation{Department of Physics, College of Science, Charmo University, 46023, Chamchamal, Sulaymaniyah, Iraq}
\author{Yidian Chen}
\email{chenyidian@hznu.edu.cn}
\affiliation{School of Physics, Hangzhou Normal University, Hangzhou, 311121, P.R. China}

\author{Mei Huang}
\email{huangmei@ucas.edu.cn}
\affiliation{School of Nuclear Science and Technology, University of Chinese Academy of Sciences, Beijing, 100049, P.R. China}

\begin{abstract}

In this study, we employ a modified soft-wall holographic model with four flavors to investigate the meson spectra, decay constants, electromagnetic form factors, and charge radius of various mesons. We obtain the spectra for vector, axial vector, and pseudoscalar mesons. Decay constants are calculated and compared with experimental and lattice QCD data. The pion and kaon electromagnetic form factors are compared with the experimental data, and a good agreement is achieved for the kaon at low $Q^{2}$. For the charmed mesons, the electromagnetic form factors of the $D$ and $D_{s}$ and electric form factors of the $D^{*}$ and $D_{s}^{*}$ are well consistent with the lattice QCD data. Moreover, the electric, magnetic, and quadrupole form factors are predicted for the $\rho$, $K^{*}$, $a_1$, $K_1$, $D_1$, and  $D_{s1}$ mesons. Furthermore, the charge radius of the vector, axial vector, and pseudoscalars, including the strange and charmed mesons, are computed. 

\end{abstract}

\maketitle

\section{Introduction}

Understanding the internal structure of hadrons involves studying electromagnetic form factors, which provide information about their magnetic moments and charge distributions. Since the electromagnetic form factors are nonperturbative quantities, one can not use the quantum chromodynamics (QCD)  to compute them. Experimental efforts have been underway for several decades to measure the electromagnetic form factors of light pseudoscalar mesons such as the pion and kaon. The pion and kaon form factors and their charge radius have been measured in Refs. \cite{NA7:1986vav,JeffersonLab:2008jve} and Ref. \cite{Amendolia:1986ui}, respectively. However, the experimental data for heavy mesons is not yet available. From  theoretical perspective, various nonperturbative approaches have been proposed in the literature, e.g., see details in Refs. \cite{Aliev:2004uj,Braguta:2005tu,Xu:2019ilh,Cheng:2019ruz,Jin:2012gu,Raha:2008ve,Stamen:2022uqh,Hwang:2009qz,Cheng:2003sm,Verma:2011yw,Hwang:2001th,Yu:2007hp}. Moreover, lattice QCD approach has extensively studied the electromagnetic form factors and charge radii of both light and heavy mesons \cite{Can:2012tx,Cui:2019rid,JLQCD:2008kdb,Frezzotti:2008dr,Bonnet:2004fr,Li:2017eic}.

The holographic QCD approach as a low energy QCD model was established in the works of Refs. \cite{Erlich:2005qh,DaRold:2005mxj,Karch:2006pv,Gubser:2008ny,Gubser:2008yx,Grefa:2021qvt,Gursoy:2007cb,Gursoy:2007er,Chen:2022goa,Li:2012ay,Li:2013oda,MartinContreras:2021yfz,Contreras:2021epz,Cao:2022csq,Chen:2022pgo} for the light mesons with the features of the dynamical chiral symmetry breaking. Extensions of this model have been made to include three flavors and flavor symmetry breaking due to the strange quark mass \cite{Abidin:2009aj, Li:2016smq}. Recent developments have further extended the holographic QCD model to four flavors \cite{Ballon-Bayona:2017bwk, Momeni:2020bmy, Chen:2021wzj}. In Refs. \cite{Ballon-Bayona:2017bwk, Momeni:2020bmy}, the masses of the $\rho$, $\omega$, and $J/\psi$ mesons are the same due to the fact that the mass term is zero in the 5D action. However, the masses of axial vector and pseudoscalar mesons differ due to non-zero mass terms. In Ref. \cite{Chen:2021wzj}, an auxiliary scalar field was introduced to the action to explicitly break the $SU(4)_{V}$ symmetry to $SU(3)_{V}$, resulting in different mass values for the $J/\psi$ meson.

The electromagnetic form factors have been investigated in the bottom-up holographic QCD model since the establishment of the model. For the first time, the pion electromagnetic form factor was studied in the hard-wall model in Ref. \cite{Grigoryan:2007wn}. Similarly, in the work of \cite{Kwee:2007dd}, the pion electromagnetic form factor discussed in the soft-wall model, and then they improved their previous results in Ref. \cite{Kwee:2007nq} by considering a background field that interpolates between the hard-wall and soft-wall models. Furthermore, the kaon form factor calculated within the hard wall-model \cite{Sang:2010kc,Abidin:2019xwu}, yielding results that align with experimental data within the uncertainties at the low $Q^{2}$. The pioneering work on the vector sector has been investigated in Refs. \cite{Grigoryan:2007vg} and \cite{Grigoryan:2007my}, where the $\rho$ meson form factor and charge radius have been studied in the hard- and soft-wall model, respectively. By taking into account the flavor symmetry breaking, the electromagnetic form factor of the $\pi$, $\rho$, $K$, $K^*$, $D$, and $D^{*}$ mesons investigated in the work of Ref. \cite{Ballon-Bayona:2017bwk}. 

In the present paper, we follow the work of Ref. \cite{Chen:2021wzj} and take the bottom-up holographic approach. Moreover, we include the higher order potential in the 5D action. Below, we proceed by calculating the masses and decay constants of the  $\pi$, $K$, $\eta$, $D$, $D_{s}$, $\rho$, $K^*$, $\omega$, $D^{*}$, $D_{s}^{*}$, $a_{1}$, $K_{1}$, $f_{1}$, $D_{1}$, and $D_{s1}$ mesons in the ground and excited states. Additionally, we predicted and compared the electromagnetic form factor of the pion and kaon mesons with the experimental data, and $D$ and $D_{s}$ mesons with the lattice QCD data. Moreover, for the vector mesons the electric, magnetic, and quadrupole form factors and charge radii of the $\rho$ and $K^{*}$ mesons predicted, then we compare the lattice results of  $D^{*}$ and $D_{S}^{*}$ mesons, we obtain a good agreement. Finally, we predict the form factors and the charge radii of the axial vector mesons, $a_{1}$, $K_{1}$, $D_{1}$, and $D_{s1}$.

The paper is organized as follows. In section \ref{modelrevisit}, we revisit the formalism of the bottom-up holographic QCD model for $N_f=4$ flavor. Section \ref{eqofmot} describes the equations of motion for vector, axial vector, and pseudoscalar fields derived from the second-order 5D action. Section \ref{couplingform} presents the derivation of coupling constants, electromagnetic form factors, and charge radii from the three-point functions obtained from the cubic-order 5D action. The numerical results are presented in section \ref{result}. Finally, we conclude our work in section \ref{conclusion}.

\section{The holographic QCD model for $N_{f}=4$}
\label{modelrevisit}

In this section, we revisit the formalism of the bottom-up holographic QCD model, specifically focusing on the inclusion of the charm quark. The original formalism for the light flavor sector ($N_f = 2$) can be found in Refs. \cite{Erlich:2005qh,Karch:2006pv} for the hard-wall and soft-wall approaches, respectively. Additional works \cite{Abidin:2009aj,Ballon-Bayona:2017bwk, Momeni:2020bmy, Chen:2021wzj} discuss the extension of the model to include strange and charm quarks. Here, we mainly follow the conventions used in Ref. \cite{Chen:2021wzj}. The five-dimensional metric defined in the Anti-de Sitter (AdS) space is given by, 
\begin{equation}
ds^{2}= \frac{1}{z^{2}} \left( \eta_{\mu \nu} dx^{\mu} dx^{\nu} + dz^{2}\right),
\end{equation}
where $\eta_{\mu \nu}$ is the four-dimensional metric in the Minkowski space, and $z$ is the fifth dimension of the AdS space, which has an inverse energy scale. It is well-known that in the AdS/CFT model for each operator in the 4D theory (also called boundary theory), there is a corresponding field in the 5D theory (bulk theory). The operators that are important for the chiral dynamics are left- and right-handed currents $J_{R/L \mu}^a=\bar{\psi}_{q R/L} \gamma_\mu t^a \psi_{q R/L}$ and quark bilinear $\bar{\psi}_{q L} \psi_{q R}$ which correspond to the $R_{\mu}^{a}$ and $L_{\mu}^{a}$ gauge fields and complex scalar fields $X$, respectively. Moreover, in Ref. \cite{Chen:2021wzj} a heavy scalar field $H$ and vector field $V_{H}$ were introduced to describe the difference between the mass of the light vector meson ($\rho$) and heavy charmed meson ($J/\psi$). Now, we have all the ingredients to write down the general five-dimensional action with $SU(4)_{L} \times SU(4)_{R}$ symmetry as follows.

\begin{equation}
\begin{aligned}
S_{M} &=-\int_{\epsilon}^{z_{m}} d^{5} x \sqrt{-g} e^{-\phi} \operatorname{Tr}\left\{\left(D^{M} X\right)^{\dagger}\left(D_{M} X\right) + M_{5}^{2}|X|^{2}\right.\\
&\left.+\frac{1}{4 g_{5}^{2}}\left(L^{M N} L_{M N}+R^{M N} R_{M N}\right)+\left(D^{M} H\right)^{\dagger}\left(D_{M} H\right) + M_{5}^{2} |H|^{2}\right\},
\end{aligned}
\label{action}
\end{equation}
where $D^{M} X=\partial_{M} X - i L_{M} X + i X R_{M}$ and $D^{M} H = \partial_{M} H - i V_{H  M} + i H V_{H  M} $ are the covariant derivative of the scalar field $X$ and $H$, respectively, and $M_{5}^{2}=(\Delta - p)(\Delta + p -4) =-3$ where $\Delta=3$ and $p=0$. Despite of introducing the dilaton field $\phi$, a hard cutoff $z_{m}$ is inserted as the upper limit of the model. So now, the limits of the integration represent the IR ($z_{m}$) and UV ($\epsilon$) regions. The coupling constant $g_{5}$ is related to the number of color ($N_{c}$) using the AdS/CFT dictionary $g_{5}^{2} = 12 \pi^{2} /N_{c}$ \cite{Erlich:2005qh}, where $N_{c}=3$. The gauge field strength $L_{M N}$ and $R_{M N}$ are defined by 

\begin{equation}
\begin{aligned}
&L_{M N}=\partial_{M} L_{N}-\partial_{N} L_{M}-i\left[L_{M}, L_{N}\right], \\
&R_{M N}=\partial_{M} R_{N}-\partial_{N} R_{M}-i\left[R_{M}, R_{N}\right].
\end{aligned}
\end{equation}

To simplify the analysis, the left- and right-handed gauge fields are replaced with the vector ($V_{M}=\frac{1}{2}(L_{M} + R_{M})$) and axial fields (($A_{M}=\frac{1}{2}(L_{M} - R_{M})$)). The decomposition allows us to express the chiral gauge fields and the covariant derivative of the complex scalar field in terms of the vector and axial fields,

\begin{equation}
\begin{aligned}
&V_{M N}=\partial_{M} V_{N}-\partial_{N} V_{M}-i\left[V_{M}, V_{N}\right] -i\left[A_{M}, A_{N}\right], \\
&A_{M N}=\partial_{M} A_{N}-\partial_{N} A_{M}-i\left[V_{M}, A_{N}\right] -i\left[A_{M}, V_{N}\right], \\
&D^{M} X=\partial_{M} X - i \left[V_{M}, X\right] - i \{A_{M}, X\} .
\end{aligned}
\end{equation}

The complex scalar field can be decomposed as 
\begin{equation}
X=e^{i \pi} X_{0} e^{i \pi}
\end{equation}
where $X_{0}=\frac{1}{2}\operatorname{diag}\left[v_{u}(z), v_{d}(z), v_{s}(z), v_{c}(z)\right]$ the vacuum expectation value, and $\pi$ is the pseudoscalar field. Moreover, The action also includes a dilaton field $\phi$. In the original hard-wall model \cite{Erlich:2005qh}, it is impossible to get the mesons' linear mass trajectories. To overcome the issue, a quadratic dilaton field introduced in Ref. \cite{Karch:2006pv}, which only depends on the fifth dimension $z$,

\begin{equation}
    \phi(z\to \infty) = \mu^{2} z^{2},
\end{equation}
where $\mu$ is related to the Regge slope and sets the mass scale for the meson spectrum.

To study the behavior of the background $X_{0}$ and find the vacuum expectation value,  one needs to turn off all the fields in the action (Eq. \eqref{action})  except the background. The equation of motion for the scalar vacuum expectation value $v_{q=l,s,c}(z)$ with $l=u,d$ is obtained as

\begin{equation}
- \frac{z^3}{e^{-\phi}} \partial_{z} \frac{e^{-\phi}}{z^3} \partial_{z} v_{q}(z) - \frac{3}{z^2} v_{q}(z)=0.
\label{vevem}
\end{equation}

The analytical solutions of Eq. \eqref{vevem} is given by the Tricomi confluent hypergeometric function $U(a,b,y)$ and the generalized Laguerre polynomial $L$.

\begin{equation}
    v_{q}(z)=  C_{q 1} z \sqrt{\pi} U(\frac{1}{2},0,\phi) - C_{q 2} z L(-\frac{1}{2},-1,\phi)
    \label{scalareom}
\end{equation}

As reported in \cite{Colangelo:2008us}, the second part of the solution must be disregarded to get a finite action at the IR region. In the UV region, Eq. \eqref{scalareom} expands to 

\begin{equation}
\begin{aligned}
     \left.v_q(z)\right|_{z \rightarrow 0} & = 2 C_{q 1} z + \\ & C_{q 1}\left[-\mu^2+2 \gamma_E \mu^2+2 \mu^2 \log z+2 \mu^2 \log \mu+\mu^2 \Psi\left(\frac{3}{2}\right)\right] z^3.
    \end{aligned}
\end{equation}

Comparing this solution to the UV behavior of the chiral condensate in the hard-wall model ($v_q(z)= m_q z + \sigma_{q} z^3$)\cite{Erlich:2005qh}, one can notice that the quark mass is $m_q = 2 C_{q 1}$ and the quark condensate $\sigma_{q}$ is also proportional with $C_{q 1}$. Now, spontaneous and explicit chiral symmetry breaking is related. In contradiction with QCD, in the massless limit, the spontaneous chiral symmetry vanishes\cite{Gherghetta:2009ac}. As pointed out in Ref. \cite{Karch:2006pv}, the issue is present because the equation of motion is linear in $v_q(z)$. A modified dilaton profile was proposed in \cite{Chelabi:2015cwn}, and this work also consider higher order potential to solve this issue
by adding  higher-order terms in the scalar potential can remedy it \cite{Karch:2006pv}. 

In Ref. \cite{Gherghetta:2009ac}, a higher order term (quartic term) in the potential $V (X)$ added to the $5D$ action,
\begin{equation}
\begin{aligned}
S_{M} &=-\int_{\epsilon}^{z_{m}} d^{5} x \sqrt{-g} e^{-\phi} \operatorname{Tr}\left\{\left(D^{M} X\right)^{\dagger}\left(D_{M} X\right)+ M_{5}^{2}|X|^{2} -\kappa |X|^4\right.\\
&\left.+\frac{1}{2 g_{5}^{2}}\left(V^{M N} V_{M N}+A^{M N} A_{M N}\right)+\left(D^{M} H\right)^{\dagger}\left(D_{M} H\right)+ M_{5}^{2}|H|^{2}\right\},
\end{aligned}
\label{actionnew}
\end{equation}
where $\kappa$ is a parameter and can be determined. Keeping only the background in the action, the zeroth order of the action is given by
\begin{equation}
\begin{aligned}
S^{(0)}=&- \frac{1}{4} \int_{\epsilon}^{z_{m}} d^{5} x\left\{\frac{e^{-\phi(z)}}{z^{3}}\left(2 v_{l}^{\prime}(z) v_{l}^{\prime}(z)+v_{s}^{\prime}(z) v_{s}^{\prime}(z)+v_{c}^{\prime}(z) v_{c}^{\prime}(z)\right) - \right.\\
&\left.\frac{e^{-\phi(z)}}{z^{5}} \left(3 \left(2 v_{l}(z)^{2}+v_{s}(z)^{2}+v_{c}(z)^{2}\right)- \frac{\kappa}{4} \left(2 v_{l}(z)^{4}+v_{s}(z)^{4}+v_{c}(z)^{4}\right) \right)+ \right.\\
&\left.\frac{e^{-\phi(z)}}{z^{3}}\left(h_{c}^{\prime}(z) h_{c}^{\prime}(z)\right) - \frac{3 e^{-\phi(z)}}{z^{5}}  h_{c}(z)^{2}\right\}
\end{aligned}
\label{zeroorder}
\end{equation}

Now, the equation of motion for the scalar vacuum expectation value is not linear anymore and becomes, 
\begin{equation}
- \frac{z^3}{e^{-\phi}} \partial_{z} \frac{e^{-\phi}}{z^3} \partial_{z} v_{q}(z) - \frac{3}{z^2} v_{q}(z) - \frac{\kappa}{2 z^2} v_{q}^{3}(z)=0
\label{eomvz}
\end{equation}

An appropriate parameterization that respects the UV and IR asymptotic behavior of the vacuum expectation value is given in Ref.  \cite{Gherghetta:2009ac}
\begin{equation}  
 v(z)=a z + b z \tanh \left(c z^2\right),
 \label{vz}
\end{equation}
with the definitions for parameters $a$, $b$, and $c$ as 
\begin{equation}
a=\frac{\sqrt{3} m_q}{g_5 }, \quad b=\sqrt{\frac{4 \mu^{2}}{\kappa}}-a, \quad c=\frac{g_5 \sigma}{\sqrt{3} b}.
\end{equation}
Now $v(z)$ in Eq. \eqref{vz} satisfies the UV and IR behavior at small and large z, 
\begin{eqnarray}
  v(z \to 0) &=& a z +  b c z^{3} + \mathcal{O}(z^{5}), \\
  v(z \to \infty) &= &(a +  b) z = \sqrt{\frac{4 \mu^{2}}{\kappa}} z.
\end{eqnarray}

Furthermore, one can obtain the dilaton profile by substituting Eq. \eqref{vz} into \eqref{eomvz} and solve the following equation for $\phi(z)$,

\begin{equation}
   \phi^{\prime}(z)= \frac{1}{\partial_{z} v_{q}(z)} \left[ z^{3} \partial_{z} \frac{1}{z^3} \partial_{z} v_{q}(z) + \frac{3}{z^2} v_{q}(z) + \frac{\kappa}{2 z^2} v_{q}^{3}(z)    \right],
\label{phifield}
\end{equation}
where the asymptotic behaviour of the dilaton profile obtained as the following,

\begin{eqnarray}
  \phi(z \to 0) &=& \frac{\kappa}{4} a^2 z^2 +  \mathcal{O}(z^{6}), \label{UVphi} \\
  \phi(z \to \infty) &=& \frac{\kappa}{4} (a+b)^2 z^2= \mu^{2} z^{2}. \label{IRphi}
\end{eqnarray}

It is worth pointing out that, the dilaton profile in Eq. \eqref{phifield} depends on the quark flavor, and for each flavor, one can obtain different dilaton profile. This behavior can be clearly seen in the UV asymptotic behaviour of $\phi(z \to 0) $ in Eq. \eqref{UVphi}, where $\phi$ is proportional to the square of the quark mass. Conversely, the IR behaviour of the $\phi$ field is similar to the one used in the original soft wall model \cite{Karch:2006pv} and does not depend on the quark mass. This IR behaviour secure the linear trajectories of the mass spectra. In order to avoid the flavor dependence of the dilaton profile and guarantee the linear Regge slope, we use the IR asymptotic behavior of the $\phi$ field in the present work. It is also noticed that comparing with the modified dilaton profile proposed in \cite{Chelabi:2015cwn}, where a negative quadratic dilaton at UV and a positive quadratic dilaton at IR, our solution obtained in Eqs.(\ref{UVphi}) and (\ref{IRphi}) requires positive quadratic dilaton at both UV and IR.

Similar to the bilinear field $X$, the auxiliary field $H$ is a diagonal matrix. However, it only reflects the effect of the charm quark mass, $H=\frac{1}{2}\operatorname{diag}\left[0, 0, 0,h_{c}(z)\right]$. Then, $h_c$ at the UV boundary should behave like $ h_{c}(z\rightarrow \infty)=m_{hc} z $. To get such a behavior, we assume that $h_c=a z$, where the value of $m_{hc}$ is different from the charm quark mass used in Eq. \eqref{vz}.

\section{Equations of motion}
\label{eqofmot}

In AdS/CFT model, one can find the mass eigenvalues by solving the equation of motion of each particular field. The equation of motion can be found from the expansion of the action in Eq. \eqref{actionnew} up to the second order in the fields $V_{M}$, $A_{m}$, and $\pi$. The fields $V_{M}$, $A_{m}$, and $\pi$ in Eq. \eqref{actionnew} expanded to $V_{M}^{a} t^{a}$, $A_{m}^{a} t^{a}$, and $\pi^{a} t^{a}$, respectively, where $t^{a}$, $a=1,2,...,15$ are the generators of $SU(4)$ group which satisfy $Tr(t^{a} t^{b})=\frac{1}{2} \delta^{ab}$. Now we can write the action up to the second order as the following

\begin{equation}
\begin{aligned}
S^{(2)}=&-\int d^{5} x\left\{\eta^{M N} \frac{e^{-\phi(z)}}{z^{3}}\left(\left(\partial_{M} \pi^{a}-A_{M}^{a}\right)\left(\partial_{N} \pi^{b}-A_{N}^{b}\right) M_{A}^{a b}-V_{M}^{a} V_{N}^{b} M_{V}^{a b}+V_{H  M}^{a} V_{H  N}^{b} M_{V_{H }}^{a b}\right)\right.\\
&\left.+\frac{e^{-\phi(z)}}{4 g_{5}^{2} z} \eta^{M P} \eta^{N Q}\left(V^{a}_{M N} V^{b}_{P Q}+A^{a}_{M N} A^{b}_{P Q}\right)\right\}
\end{aligned}
\label{S2}
\end{equation}
where $\eta^{M N}$ is the metric in 5-D Minkowski space which is $\eta^{M N}=diag[-1,1,1,1,1]$ , $V^{a}_{M N}= \partial_{M} V^{a}_{N} - \partial_{N} V^{a}_{M}$, and $A^{a}_{M N}= \partial_{M} A^{a}_{N} - \partial_{N} A^{a}_{M}$. The vector, axial and psudoscalar fields are described by $4 \times 4$ matrices,

\begin{equation}
\begin{aligned}
& V=V^a t^a=\frac{1}{\sqrt{2}}\left(\begin{array}{cccc}
\frac{\rho^0}{\sqrt{2}}+\frac{\omega^{\prime}}{\sqrt{6}}+\frac{\psi}{\sqrt{12}} & \rho^{+} & K^{*+} & \bar{D}^{* 0} \\
\rho^{-} & -\frac{\rho^0}{\sqrt{2}}+\frac{\omega^{\prime}}{\sqrt{6}}+\frac{\psi}{\sqrt{12}} & K^{* 0} & D^{*-} \\
K^{*-} & \bar{K}^{* 0} & -\sqrt{\frac{2}{3}} \omega^{\prime}+\frac{\psi}{\sqrt{12}} & D_s^{*-} \\
D^{* 0} & D^{*+} & D_s^{*+} & -\frac{3}{\sqrt{12}} \psi
\end{array}\right), 
\end{aligned}
\end{equation}
\begin{equation}
\begin{aligned}
& A=A^a t^a=\frac{1}{\sqrt{2}}\left(\begin{array}{cccc}
\frac{a_1^0}{\sqrt{2}}+\frac{f_1}{\sqrt{6}}+\frac{\chi_{c 1}}{\sqrt{12}} & a_1^{+} & K_1^{+} & \bar{D}_1^0 \\
a_1^{-} & -\frac{a_1^0}{\sqrt{2}}+\frac{f_1}{\sqrt{6}}+\frac{\chi_{c 1}}{\sqrt{12}} & K_1^0 & D_1^{-} \\
K_1^{-} & \bar{K}_1^0 & -\sqrt{\frac{2}{3}} f_1+\frac{\chi_{c 1}}{\sqrt{12}} & D_{s 1}^{-} \\
D_1^0 & D_1^{+} & D_{s 1}^{+} & -\frac{3}{\sqrt{12}} \chi_{c 1}
\end{array}\right),
\end{aligned}
\end{equation}
\begin{equation}
\begin{aligned}
& \pi=\pi^a t^a=\frac{1}{\sqrt{2}}\left(\begin{array}{cccc}
\frac{\pi^0}{\sqrt{2}}+\frac{\eta}{\sqrt{6}}+\frac{\eta_c}{\sqrt{12}} & \pi^{+} & K^{+} & \bar{D}^0 \\
\pi^{-} & -\frac{\pi^0}{\sqrt{2}}+\frac{\eta}{\sqrt{6}}+\frac{\eta_c}{\sqrt{12}} & K^0 & D^{-} \\
K^{-} & \bar{K}^0 & -\sqrt{\frac{2}{3}} \eta+\frac{\eta_c}{\sqrt{12}} & D_s^{-} \\
D^0 & D^{+} & D_s^{+} & -\frac{3}{\sqrt{12}} \eta_c
\end{array}\right) .
\end{aligned}
\end{equation}

The mass terms in the action $M_{A}^{a b}$, $M_{V}^{a b}$, and $M_{V_{H }}^{a b}$ are defined by

\begin{equation}
\begin{aligned}
&M_{A}^{a b} \delta^{a b}= Tr\left( \{t^{a},X_{0}\} \{t^{b},X_{0}\} \right), \\
&M_{V}^{a b} \delta^{a b}= Tr\left( [t^{a},X_{0}] [t^{b},X_{0}] \right), \\
&M_{V_{H }}^{a b} = Tr\left( [H,t^{a}] [H,t^{b}] \right), \\
\end{aligned}
\end{equation}
where $M_{V_{H }}^{a,b}$ is zero except for the case $a=b=15$. The vector field in Eq. \eqref{S2} satisfies the following equation of motion,

\begin{equation}
- \partial^{M} \frac{e^{-\phi}}{g_{5}^{2} z} V_{M N}^{a} - \frac{e^{-\phi}}{ z^{3}} \left( M_{V}^{a a} V_{M}^{a} -  M_{V_{H }}^{a a} V_{H  M}^{a}\right)=0,
\label{emVm}
\end{equation}
where $V_{M}^{a}=(V_{\mu}^{a}, V_{z}^{a})$. The gauge fixing for the vector field is $V_{z}^{a}=0$ and for the transverse part of the vector field $( V_{\mu}^{a} = V_{\mu \perp}^{a} + V_{\mu \parallel}^{a})$, $\partial^{\mu} V_{\mu \perp}^{a}=0$. Applying the decomposition and gauge choices, Eq. \eqref{emVm} reduces to the following form
\begin{equation}
\left(-\frac{z}{e^{-\phi}} \partial_{z} \frac{e^{-\phi}}{z} \partial_{z}+\frac{2 g_{5}^{2}\left(M_{V_{H }}^{a a}-M_{V}^{a a}\right)}{z^{2}}\right) V_{\mu \perp}^{a}(q, z)=-q^{2} V_{\mu \perp}^{a}(q, z),
\label{eqV}
\end{equation}
where $V_{\mu \perp}^{a}(q, z)$ is the 4D Fourier transform of $V_{\mu \perp}^{a}(x, z)$. According to AdS/CFT conjecture, one can write the vector field according to the bulk-to-boundary propagator and its boundary value at UV, which acts as a Fourier transform of the source of the 4D conserved vector current operator, $V_{\mu \perp}^{a}(q, z)= V_{\mu \perp}^{0a}(q) \mathcal{V}^{a}(q^{2},z)$. It is worth knowing that the bulk-to-boundary propagator $\mathcal{V}^{0a}(q^{2},z)$ satisfies the equation of motion for the vector field with the boundary condition $\mathcal{V}^{a}(q^{2},\epsilon)=1$ and $\partial_{z} \mathcal{V}^{a}(q^{2},zm)=0$. To find the mass eigenvalues of the vector mesons, we need a wave-function $\psi_{V}(z)$ with $-q^{2}=m_{V}^{2}$, which satisfies Eq. \eqref{eqV} with the boundary conditions $\psi_{V^{n}}(\epsilon)=0$ and $\partial_{z} \psi_{V^{n}}(zm)=0$, and normalized as $ \int dz \frac{e^{-\phi}}{z} \psi^{n}_{V}(z) \psi^{m}_{V}(z)=\delta^{n m}$. Moreover, the bulk-to-boundary propagator can be written as a sum over meson poles.

\begin{equation}
    \mathcal{V}^{a}(q^{2},z)= \sum_{n} \frac{- g_{5} f_{V^{n}}^{a} \psi_{V^{n}}^{a}(z)}{q^{2} - m_{V^{n}}^{a^{2}}},
\label{bulktoboundary}
\end{equation}
where $f_{V^{n}}^{a}= |\partial_{z} \psi_{V^{n}}^{a}(\epsilon)/(g_{5} \epsilon) |$ is the decay constant of the $n^{th}$ mode of the vector meson \cite{Erlich:2005qh}.

The equation of motion for the axial vector field is similar to the vector field, except the auxiliary field $H$ does not contribute. The axial vector field $A_{\mu}^{a}$ can be decomposed to the transverse and longitudinal parts, $A_{\mu}^{a}= A_{\mu \perp}^{a} + A_{\mu \parallel}^{a}$, where $A_{\mu \parallel}^{a}=\partial_{\mu}\phi^{a}$ has the contribution to the pesudoscalar mesons. The equation of motion derived from Eq. \eqref{actionnew} is given by

\begin{equation}
\left(-\frac{z}{e^{-\phi}} \partial_{z} \frac{e^{-\phi}}{z} \partial_{z}+\frac{2 g_{5}^{2}M_{A}^{a a}}{z^{2}}\right) A_{\mu \perp}^{a}(q, z)=-q^{2} A_{\mu \perp}^{a}(q, z),
\label{eqA}
\end{equation}
with the gauge fixing and transverse condition $A_{z}^{a}=0$, and $\partial^{\mu} A_{\mu \perp}^{a}$, respectively. The bulk-to-boundary propagator of the axial vector field $\mathcal{A}^{a}(q^{2},z)$ satisfy the bounadry conditions $\mathcal{A}^{a}(q^{2},\epsilon)=0$ and $\partial_{z} \mathcal{A}^{a}(q^{2},zm)=0$, in the UV and IR region. 

Finally, the mass spectra of the pseudoscalar mesons can be found by solving the coupled equation of motions between the pseudoscalar field $\pi$ and the longitudinal part of the axial vector field $\phi$,

\begin{equation}
\begin{aligned}
& q^2 \partial_z \varphi^a(q, z)+\frac{2 g_5^2 M_A^{a a}}{z^2} \partial_z \pi^a(q, z)=0, \\
& \frac{z}{e^{-\phi}} \partial_z\left(\frac{e^{-\phi}}{z} \partial_z \varphi^a(q, z)\right)-\frac{2 g_5^2 M_A^{a a}}{z^2}\left(\varphi^a(q, z)-\pi^a(q, z)\right)=0,
\end{aligned}
\label{empesudo}
\end{equation}
with the boundary conditions $\pi^{a}(q^{2},\epsilon)=\phi^{a}(q^{2},\epsilon)=0$ and $\partial_{z}\pi^{a}(q^{2},zm)=\partial_{z}\phi^{a}(q^{2},zm)=0$. In the holographic QCD approach, the decay constants for the vector, axial vector, and pseudoscalar mesons are defined by

\begin{equation}
\begin{aligned}
&f_{V^{n}}^{a}= \frac{e^{-\phi}}{g_{5} z} \partial_{z} \psi_{V^n}^{a}(z)|_{z\to \epsilon}, \\
&f_{A^{n}}^{a}= \frac{e^{-\phi}}{g_{5} z} \partial_{z} \psi_{A^n}^{a}(z)|_{z\to \epsilon}, \\
&f_{\pi^{n}}^{a}= -\frac{e^{-\phi}}{g_{5} z} \partial_{z} \varphi^{a,n}(z)|_{z\to \epsilon} , \\
\end{aligned}
\end{equation}
where $\psi_{V^n}^{a}(z)$, $\psi_{A^n}^{a}(z)$, and $\varphi^{a,n}(z)$ are the normalized wavefunctions for the vector, axial vector, and pseudoscalar mesons, respectively.

\section{Three-point interaction Coupling Constants and form factors}
\label{couplingform}

The coupling constants and form factors can be obtained from the cubic order of the 5D action in Eq. \eqref{actionnew},

\begin{equation}
\begin{aligned}
S^{(3)}= & -\int d^{5} x\left\{\eta^{M N} \frac{e^{-\phi(z)}}{z^{3}}(2\left(A_{M}^{a}-\partial_{M} \pi^{a}\right) V_{N}^{b} \pi^{c} g^{a b c}+V_{M}^{a}\left(\partial_{N}\left(\pi^{b} \pi^{c}\right)-2 A_{M}^{b} \pi^{c}\right) h^{a b c}\right.\\
& -  V_{M}^{a} V_{N}^{b} \pi^{c} k^{a b c}) +\frac{e^{-\phi(z)}}{2 g_{5}^{2} z} \eta^{M P} \eta^{N Q}(V_{M N}^{a} V_{P}^{b} V_{Q}^{c}+V_{M N}^{a} A_{P}^{b} A_{Q}^{c}+A_{M N}^{a} V_{P}^{b} A_{Q}^{c} 
\left. +A_{M N}^{a} A_{P}^{b} V_{Q}^{c}) f^{b c a}\right\}
\end{aligned}
\end{equation}
with the following definitions for $g^{a b c}$, $h^{a b c}$, and $k^{a b c}$,

\begin{equation}
\begin{aligned}
&g^{a b c} = i Tr\left( \{t^{a},X_{0}\} [t^{b},\{t^{c},X_{0}\}] \right), \\
&h^{a b c} = i Tr\left( [t^{a},X_{0}] \{t^{b},\{t^{c},X_{0}\}\} \right), \\
&k^{a b c} = -2 Tr\left( [t^{a},X_{0}] [t^{b},\{t^{c},X_{0}\}] \right). \\
\end{aligned}
\end{equation}

In the present work, we are interested in the three-point interactions of the $VVV$, $VAA$, and $V\pi\pi$. The corresponding part of the action to these three-point interactions are

\begin{equation}
    S_{VVV}=  -\int_{\epsilon}^{z_{m}} d^{5} \frac{e^{-\phi(z)}}{2 g_{5}^{2} z} \eta^{m p} \eta^{n q}\left(V_{m n}^{a} V_{p}^{b} V_{q}^{c}\right) f^{a b c}
\label{actionvector}
\end{equation}

\begin{equation}
    S_{VAA}=  -\int_{\epsilon}^{z_{m}} d^{5} \frac{e^{-\phi(z)}}{2 g_{5}^{2} z} \eta^{m p} \eta^{n q}\left(V_{m n}^{a} A_{p}^{b} A_{q}^{c}+A_{m n}^{a} V_{p}^{b} A_{q}^{c}+A_{m n}^{a} A_{p}^{b} V_{q}^{c}\right)  f^{a b c}
\label{actionaxial}
\end{equation}

\begin{equation}
\begin{aligned}
S_{V\pi\pi}= & -\int_{\epsilon}^{z_{m}} d^{5} x\left\{\eta^{m n} \frac{e^{-\phi(z)}}{z^{3}}\left(2\left(A_{m}^{a}-\partial_{m} \pi^{a}\right) V_{n}^{b} \pi^{c} g^{a b c}+V_{m}^{a}\left(\partial_{n}\left(\pi^{b} \pi^{c}\right)-2 A_{n}^{b} \pi^{c}\right) h^{a b c}\right)\right.\\
&\left. +\frac{e^{-\phi(z)}}{2 g_{5}^{2} z} \eta^{m p} \eta^{n q}\left(V_{m n}^{a} A_{p}^{b} A_{q}^{c}\right) f^{abc}\right\}
\end{aligned}
\label{actionpion}
\end{equation}

In the AdS/CFT approach, the coupling constants of three-point interaction can be obtained from the three-point correlation functions by functional variation of the 5D action with respect to the sources of the 5D fields \cite{Grigoryan:2007wn,Kwee:2007nq}. The relevant coupling constants calculated are given by

\begin{equation}
    g_{VVV}= g_{5} f^{abc} \int_{\epsilon}^{zm} dz  \frac{e^{-\phi(z)}}{ z} V^{a}(z) V^{b}(z) V^{c}(z),
\label{vectorcoupling}
\end{equation}

\begin{equation}
    g_{VAA}= g_{5} f^{abc} \int_{\epsilon}^{zm} dz  \frac{e^{-\phi(z)}}{ z} V^{a}(z) A^{b}(z) A^{c}(z),
\label{axialcoupling}
\end{equation}

\begin{equation}
   g_{V\pi\pi}= g_{5} \int_{\epsilon}^{zm} dz  \frac{e^{-\phi(z)}}{ z} \left( f^{abc} \partial_{z}\phi^{a} V^{b}(z) \partial_{z}\phi^{a} - \frac{2 g_{5}^{2}}{z^{2}} (\pi^{a}-\phi^{a}) V^{b}(z) (\pi^{c}-\phi^{c})  (g^{abc} - h^{bac}) \right) .
\label{pioncoupling}
\end{equation}

Moreover, we study the electromagnetic (EM) form factors of the channels $VVV$, $VAA$, and $V\pi\pi$ and the charge radius. The EM form factors of the vector, axial vector, and pseudoscalar mesons are defined by \cite{Ballon-Bayona:2017bwk,Grigoryan:2007wn,BallonBayona:2009ar}

\begin{equation}
\begin{aligned}
 \left\langle V^a(p+q), \epsilon^{\prime}\left|J_{E M}^\mu(0)\right| V^a(p), \epsilon\right\rangle  =& -\left(\epsilon^{\prime} \cdot \epsilon\right)(2 p+q)^\mu F_{V^a}^1\left(q^2\right) \\
& +\left[\epsilon^{\prime \mu}(\epsilon \cdot q)-\epsilon^\mu\left(\epsilon^{\prime} \cdot q\right)\right]\left[F_{V^a}^1\left(q^2\right)+F_{V^a}^2\left(q^2\right)\right] \\
& +\frac{1}{M_{V^a}^2}\left(q \cdot \epsilon^{\prime}\right)(q \cdot \epsilon)(2 p+q)^\mu F_{V^a}^3\left(q^2\right) .
\end{aligned}
\label{emVform}
\end{equation}

\begin{equation}
\begin{aligned}
\left\langle A^a(p+q), \epsilon^{\prime}\left|J_{E M}^\mu(0)\right| A^a(p), \epsilon\right\rangle
 =& -\left(\epsilon^{\prime} \cdot \epsilon\right)(2 p+q)^\mu F_{A^a}^1\left(q^2\right) \\
& +\left[\epsilon^{\prime \mu}(\epsilon \cdot q)-\epsilon^\mu\left(\epsilon^{\prime} \cdot q\right)\right]\left[F_{A^a}^1\left(q^2\right)+F_{A^a}^2\left(q^2\right)\right] \\
& +\frac{1}{M_{A^a}^2}\left(q \cdot \epsilon^{\prime}\right)(q \cdot \epsilon)(2 p+q)^\mu F_{A^a}^3\left(q^2\right) .
\end{aligned}
\label{emAform}
\end{equation}

\begin{equation}
\left\langle\pi^a(p+q)\left|J_{E M}^\mu(0)\right| \pi^a(p)\right\rangle=(2 p+q)^\mu F_{\pi^a}\left(q^2\right).
\end{equation}
where $\epsilon$ and $\epsilon^{\prime}$ are the polarization vectors of the initial and final vector mesons, respectively, and  $J_{E M}^\mu(0)$ is the electromagnetic current. The electromagnetic current can be represented by the linear combination of the flavor currents via \cite{Ballon-Bayona:2017bwk}

\begin{equation}
    J_{E M}^\mu(x) = \sum_{a=(\rho,\omega,J/\psi)} c_{a} J_{a}^\mu(x)
\end{equation}
where the coefficients $c_{a}$ is a constant depending on the contribution of the quarks in the EM current. The electric (charge), magnetic, and quadrupole form factors of the vector and axial vector mesons can be deduced from the linear combination of the form factors in Eq. \eqref{emVform} and Eq. \eqref{emAform}, respectively.

\begin{equation}
\begin{aligned}
&F_{V^{a}}^{E}= F_{V^a}^1 + \frac{q^{2}}{6 M_{V^{a}}^{2}} \left[F_{V^a}^2 - \left(1 - \frac{q^{2}}{4 M_{V^{a}}^{2}}  \right) F_{V^a}^3 \right] , \\
&F_{V^a}^M = F_{V^a}^1 + F_{V^a}^2, \\
&F_{V^{a}}^{Q}= - F_{V^a}^2 +  \left(1 - \frac{q^{2}}{4 M_{V^{a}}^{2}}  \right) F_{V^a}^3 . \\
\end{aligned}
\end{equation}

\begin{equation}
\begin{aligned}
&F_{A^{a}}^{E}= F_{A^a}^1 + \frac{q^{2}}{6 M_{A^{a}}^{2}} \left[F_{A^a}^2 - \left(1 - \frac{q^{2}}{4 M_{A^{a}}^{2}}  \right) F_{A^a}^3 \right] , \\
&F_{A^a}^M = F_{A^a}^1 + F_{A^a}^2, \\
&F_{A^{a}}^{Q}= - F_{A^a}^2 +  \left(1 - \frac{q^{2}}{4 M_{A^{a}}^{2}}  \right) F_{A^a}^3 . \\
\end{aligned}
\end{equation}

For the elastic case, the electromagnetic form factors obtained as 

\begin{equation}
\begin{aligned}
&F_{V^a}^1= F_{V^a}^2 =F_{V^a}= \sum_{n} \frac{f_{V^{a}} g_{VVV}}{M_{V^{a}}^{2} + Q^{2}}  , \\
&F_{V^a}^3 = 0 ,\\
\end{aligned}
\end{equation}

\begin{equation}
\begin{aligned}
&F_{A^a}^1= F_{A^a}^2 =F_{A^a}= \sum_{n} \frac{f_{V^{a}} g_{VAA}}{M_{V^{a}}^{2} + Q^{2}}  , \\
&F_{A^a}^3 = 0 ,\\
\end{aligned}
\end{equation}

\begin{equation}
    F_{\pi^{a}}= \sum_{n} \frac{f_{V^{a}} g_{V\pi\pi}}{M_{V^{a}}^{2} + Q^{2}}
\end{equation}
where $Q^{2}=- q^{2}$. Using the couplings in Eqs. (\ref{vectorcoupling}-\ref{pioncoupling}) and the definition of the bulk-to-boundary in Eq. \eqref{bulktoboundary}, one can reach the final version of the form factors for the vector, axial vector, and pseudoscalar mesons,

\begin{equation}
    F_{V}(Q^{2})=  f^{abc} \int dz  \frac{e^{-\phi(z)}}{ z} \mathcal{V}^{a}(Q^{2},z) V^{b}(z) V^{c}(z),
\label{VVVform}
\end{equation}

\begin{equation}
    F_{A}(Q^{2})=  f^{abc} \int dz  \frac{e^{-\phi(z)}}{ z} \mathcal{V}^{a}(Q^{2},z) A^{b}(z) A^{c}(z),
\label{VVAform}
\end{equation}

\begin{equation}
   F_{\pi} (Q^{2})=  \int dz  \frac{e^{-\phi(z)}}{ z} \mathcal{V}^{b}(Q^{2},z) \left( f^{abc} \partial_{z}\phi^{a}  \partial_{z}\phi^{c} - \frac{2 g_{5}^{2}}{z^{2}} (\pi^{a}-\phi^{a})  (\pi^{c}-\phi^{c})  (g^{abc} - h^{bac}) \right) .
\label{Vpipiform}
\end{equation}

Note that the bulk-to-boundary propagator for vector sector $\mathcal{V}^{b}(Q^{2},z)$ can be obtained by solving the equation of motion Eq. \eqref{eqV}. The analytical solution of Eq. \eqref{eqV} can be obtained only for the off-shell mode of the $\rho$ meson; for others, one has to use the numerical method. The solution reads

\begin{equation}
    \mathcal{V}^{b}(Q^{2},z)= \Gamma(1+\frac{Q^{2}}{4 \mu^2}) U(\frac{Q^{2}}{4 \mu^2}, 0 , z^2 \mu^2)
\end{equation}

The charge radius can be obtained from the low $Q^{2}$ expansion of the EM form factors of a pseudoscalar meson and the charge form factors of the vector meson,

\begin{equation}
    F^{(C)}(Q^{2})= 1- \frac{1}{6} \left<r^{2}\right> Q^{2} + ...,
\end{equation}
then, the charge radius is given by

\begin{equation}
    \left<r^{2}\right> = - 6 \frac{dF(Q^{2})}{dQ^{2}}|_{Q^2\rightarrow 0 }.
\label{chargeradii}
\end{equation}

\section{Results}
\label{result}

This section presents the numerical results for the meson spectra, decay constants, electromagnetic form factor, and charge radii obtained from our holographic QCD model. The model incorporates various free parameters, such as $\mu$, $m_{u}$, $m_{s}$, $m_{c}$, $m_{hc}$, $\sigma_u$, $\sigma_s$, $\sigma_c$, $z_{m}$ and $\kappa$. The values of these parameters are determined by fitting the model with the experimental value of the mass and decay constant of the following mesons, $m_{\rho}$, $m_{\pi}$, $f_{\pi}$, $m_{a_1}$, $m_{K}$, $f_{K}$, $m_{\eta_{c}}$, $m_{\chi_{c1}}$, $m_{J/\psi}$, and $m_{\psi(3770)}$. The values of the parameters obtained from the fitting are provided in Table \ref{tab:freepar}. 

\subsection{Meson spectra}
Once the parameters are fixed, the model enables the calculation of meson spectra and decay constants for vector, axial vector, and pseudoscalar mesons. The masses of vector mesons are determined by solving the equation of motion given in Eq. \eqref{eqV}, where $m_{V}^{2}=-q^2$.  Similarly, by solving Eqs. \eqref{eqA} and \eqref{empesudo}, one can get the masses of the axial vector and pseudoscalar mesons. The resulting meson masses compared to the values listed in the Particle Data Group (PDG) \cite{ParticleDataGroup:2020ssz} are presented in Tables \ref{tab:vector}, \ref{tab:axial}, and \ref{tab:pseudoscalar} for vector, axial vector and pseudoscalar mesons, respectively. Considering the exact $SU(4)_{V}$ symmetry, the masses of the $\rho$, $\omega$, and $J/\psi$ mesons should be equal. As a consequence of the $SU(4)_{V}$ symmetry, we can see that from Table \ref{tab:vector}, the masses of the ground and excited states of the $\rho$ and $\omega$ mesons are exactly the same. However, the mass of the $J/\psi$ meson differs because of the explicit symmetry breaking of the action by introducing an auxiliary field $H$. For the case of $D^{*}$ and $D_{s}^{*}$ mesons, the inclusion of the nonzero value of $M_{V}^{aa}$ in the equation of motion (Eq. \eqref{emVm}) leads an explicit $SU(4)$ flavor symmetry breaking. In addition, for the case of the axial vector and pseudoscalar mesons, the mass term $M_{A}^{aa}$ is nonzero for all the mesons, then the chiral symmetry is explicitly broken by the different values of the quark masses for all the axial vector and pseudoscalar mesons. 
Moreover, comparing our meson spectra with the one obtained in Ref. \cite{Chen:2021wzj}, one can conclude that our results are well improved, especially for the pesudoscalar mesons. This improvement coming from the inclusion of the higher-order terms in the scalar potential.

\begin{table}
\center
\begin{tabular}{c  c  c}
\hline
\hline

     $\mu = 430$    &      &   $\sigma_u = (296.2)^3$       \\ 
     $m_u = 3.2$      &      &   $\sigma_s = (259.8)^3$    \\ 
     $m_s =142.3$   &      &   $\sigma_c = (302)^3$    \\ 
     $m_c =1597.1$    &      &   $z_{m}=10 000$             \\ 
     $m_{hc} =1985  $   &      & $\kappa =30$            \\ 
  
     \hline
     \hline
\end{tabular}
\caption{The values of the free parameters with the unit of MeV.}
\label{tab:freepar}
\end{table}

\begin{table}
\center
 \begin{tabular}{|c|cc|cc|cc|cc|cc|cc|}
\hline
\hline
        n   & $m_{\rho}$ & Exp. & $m_{K^*}$ & Exp. & $m_{\omega}$ & Exp. & $m_{D^*}$ & Exp. & $m_{D_{s}^{*}}$ & Exp. & $m_{J/\psi}$ & Exp.\\ 
\hline
     1  &860     & 775  & 860.05 & 892  & 860    & 782 & 1914.90 &2007& 1911.40 &2112 &3099.21 & 3097   \\ 
     2  &1216.24 & 1465 & 1216.29& 1414 &1216.24 & 1410& 2110.04 &2627 & 2107.36 &2714 &3329.54 & 3686   \\ 
     3  &1490.09 & 1570 & 1490.15& 1718 &1490.09 & 1670& 2286.48&2781 &         &     &3702.95 & 3733  \\ 
     4  &1726.63 & 1720 &        &      &1726.63 & 1960&        &     &         &     &3863.62 & 4040  \\ 
     5  &1957.02 & 1900 &        &      &1957.02 & 2205&        &     &         &     &        &       \\ 
     6  &2201.44 & 2150 &        &      &2201.44 & 2290&        &     &         &     &        &       \\ 
     7  &        &      &        &      &2462.12 & 2330&        &     &         &     &        &       \\ 
     
     \hline
     \hline
\end{tabular}
\caption{ Comparison of the vector mesons masses with the values listed in PDG \cite{ParticleDataGroup:2020ssz}.}
\label{tab:vector}
\end{table}

\begin{table}
\center
\begin{tabular}{|c|cc|cc|cc|cc|cc|cc|}
\hline
\hline
        n   & $m_{a_{1}}$ & Exp. & $m_{K_1}$ & Exp. & $m_{f_1}$ & Exp. & $m_{D_1}$ & Exp. & $m_{D_{s1}}$ & Exp. & $m_{\chi_{cl}}$ & Exp.\\ 
\hline
     1  &1286.95 & 1230 & 1287.67& 1253 &1287.97 & 1282& 2641.47&2422 & 2657.55 &2460 &3511.04 & 3511   \\ 
     2  &1541.88 & 1411 & 1542.90& 1403 &1543.32 & 1426&        &     &         &     &4050.30 & 3872   \\ 
     3  &1765.28 & 1655 & 1766.30& 1672 &1766.73 & 1518&        &     &         &     &4149.46 & 4147  \\ 
     4  &1968.94 & 1930 &        &      &1970.32 & 1971&        &     &         &     &4340.88 & 4274  \\ 
     5  &2172.45 & 2096 &        &      &2173.82 & 2310&        &     &         &     &        &       \\ 
     6  &2393.35 & 2270 &        &      &        &     &        &     &         &     &        &       \\ 

     \hline
     \hline
\end{tabular}
\caption{Comparison of the axial vector mesons masses with the values listed in PDG \cite{ParticleDataGroup:2020ssz}.}
\label{tab:axial}
\end{table}

\begin{table}
\center
\begin{tabular}{|c|cc|cc|cc|cc|cc|cc|}
\hline
\hline
        n   & $m_{\pi^0}$ & Exp. & $m_{K^0}$ & Exp. & $m_{\eta}$ & Exp. & $m_{D^0}$ & Exp. & $m_{D_{s}^{\pm}}$ & Exp. & $m_{\eta_{c}}$ & Exp.\\ 
\hline
     1  &141.69  & 139.57& 622.15& 498 &740.52  & 548 & 2032.67&1865 & 2114.32 &1968 &2968.46 & 2984  \\ 
     2  &1439.04 & 1300 & 1451.28& 1482 &145.78  & 1294& 2913.25&2549 &         &     &3894.14 & 3637  \\ 
     3  &1698.20 & 1810 & 1709.18& 1629 &1716.10 & 1475&        &     &         &     &        &       \\ 
     4  &2124.64 & 2070 & 1927.17& 1874 &1933.92 & 1751&        &     &         &     &        &       \\ 
     5  &2345.61 & 2360 &        &      &2142.68 & 2010&        &     &         &     &        &       \\ 
     
     \hline
     \hline
\end{tabular}
\caption{Comparison of the pseudoscalar mesons masses with the values listed in PDG \cite{ParticleDataGroup:2020ssz}.}
\label{tab:pseudoscalar}
\end{table}

\subsection{Decay constant}
The calculated decay constants are presented in Table \ref{tab:decay} for various mesons within our model. In the vector meson sector, we compared the decay constant of the $\rho$ meson to the experimental value obtained from Ref. \cite{Donoghue:1992dd}. Our result shows a discrepancy of $16\%$ compared to the experimental value. Additionally, we predicted the decay constants for the $K^{*}$, $D^{*}$, and $D_{s}^{*}$ mesons. Moving to the axial vector sector, only the experimental value for the $a_{1}$ meson (taken from Ref. \cite{Isgur:1988vm}) is available for the comparison, while the decay constants of the other axial vector mesons are predicted in our model. Notably, the predicted value for the $a_{1}$ meson aligns well with the experimental value. In the pseudoscalar sector, the decay constants of the pion and kaon are found to be compatible with the values listed in PDG \cite{ParticleDataGroup:2020ssz}.  However, for the $D^{0}$ and $D_{s}$ mesons, there are only results from lattice QCD studies, and no experimental data is directly comparable in this context. Similar to the meson spectra, the decay constants of our work have better compatibility with the experimental and lattice data compare to Ref. \cite{Chen:2021wzj}.

\begin{table}
\center
\begin{tabular}{ c  c c  c  c }
\hline
\hline
          Observable             &    & $N_{f}=4$, hQCD (MeV) & &  Measured (MeV)    \\
\hline
          $f_{\rho}^{1/2}$       &    &  288.50     &  &   $345$  \cite{Donoghue:1992dd}          \\ 
\hline
          $f_{K^*}^{1/2}$        &    &  288.28     &  &                \\ 
\hline
          $f_{D^*}^{1/2}$        &    &  413.36    &  &               \\ 
\hline      
          $f_{D_{s}^*}^{1/2}$    &    &  427.78    &  &                \\ 
\hline      
          $f_{a_1}^{1/2}$        &    &  351.34    &  &  $354 $  \cite{Isgur:1988vm}          \\ 
\hline      
          $f_{K_1}^{1/2}$        &    &  348.26    &  &               \\ 
\hline     
          $f_{f_1}^{1/2}$        &    &  346.79    &  &               \\ 
\hline      
          $f_{D_1}^{1/2}$        &    &  502.11    &  &               \\ 
\hline      
          $f_{D_s1}^{1/2}$       &    &  475.74    &  &               \\           
\hline      
          $f_{\pi}$              &    &  91.03     &  &  92.07  \cite{ParticleDataGroup:2020ssz}           \\ 
\hline      
          $f_{K}$                &    &  108.5    &  &  110  \cite{ParticleDataGroup:2020ssz}           \\ 
\hline      
          $f_{\eta}$             &    &  126.32    &  &               \\ 
\hline      
          $f_{D^0}$              &    &  199.31   &  &  149.8  \cite{ParticleDataGroup:2020ssz}           \\ 
\hline      
          $f_{D_s}$              &    &  197.73    &  &  176.1   \cite{ParticleDataGroup:2020ssz}          \\ 
                                          
     \hline
     \hline
\end{tabular}
\caption{The predicted decay constants calculated from the hQCD 
compared to experimental or lattice data.}
\label{tab:decay}
\end{table}

\subsection{Three-particle coupling constant}
In our analysis of three-point functions, we have calculated the coupling constants for three types of interactions: $VVV$ (vector-vector-vector), $VAA$ (vector-axial vector-axial vector), and $V\pi\pi$ (vector-pseudoscalar-pseudoscalar). It is worth noting that, assuming the equal masses and condensates for the quarks, the couplings should satisfy the following relations, which is the manifestation of flavor symmetry restoration, 

\begin{table}
\center
\begin{tabular}{ c  c c   }
\hline
\hline
          $SU(4)$ symmetry relation                &    & Violation in $N_{f}=4$, hQCD     \\
\hline
          $2 g_{\rho K K}/g_{\rho \pi \pi} $       &    &  $14\%$              \\ 
\hline
          $2 g_{\rho D D}/g_{\rho \pi \pi} $        &    &  $47\%$    \\ 
\hline
          $2 g_{\rho K^{*} K^{*}}/g_{\rho \rho \rho} $        &    &  $0\%$   \\ 
\hline      
          $2 g_{\rho D^{*} D^{*}}/g_{\rho \rho \rho} $    &    &  $36\%$     \\ 
\hline      
          $2 g_{\rho K_{1} K_{1}}/g_{\rho a_{1} a_{1}} $        &    &  $1\%$         \\ 
\hline      
          $2 g_{\rho D_{1} D_{1}}/g_{\rho a_{1} a_{1}} $        &    &  $76\%$   \\ 
                                          
     \hline
     \hline
\end{tabular}
\caption{ Reporting the flavor symmetry breaking in our calculations.}
\label{tab:couplingratio}
\end{table}

\begin{equation}
\begin{aligned}
&g_{\rho K K}= g_{\rho D D} = \frac{1}{2} g_{\rho \pi \pi},\\
&g_{\rho K^{*} K^{*}}= g_{\rho D^{*} D^{*}} = \frac{1}{2} g_{\rho \rho \rho},\\
&g_{\rho K_{1} K_{1}}= g_{\rho D_{1} D_{1}} = \frac{1}{2} g_{\rho a_{1}. a_{1}}
\end{aligned}
\end{equation}

The numerical value of the ratios between the strong coupling constants are reported in Table \ref{tab:couplingratio}. The $SU(4)$ flavor symmetry is clearly seen in Table \ref{tab:couplingratio}, where the symmetry is badly broken in the case of including charm quark. Similar results are reported in the hard-wall model \cite{Ballon-Bayona:2017bwk} and QCD sum rules \cite{Bracco:2011pg}. 

\subsection{Form factor}
\begin{figure}
  \centering
  \includegraphics[width=0.45\linewidth]{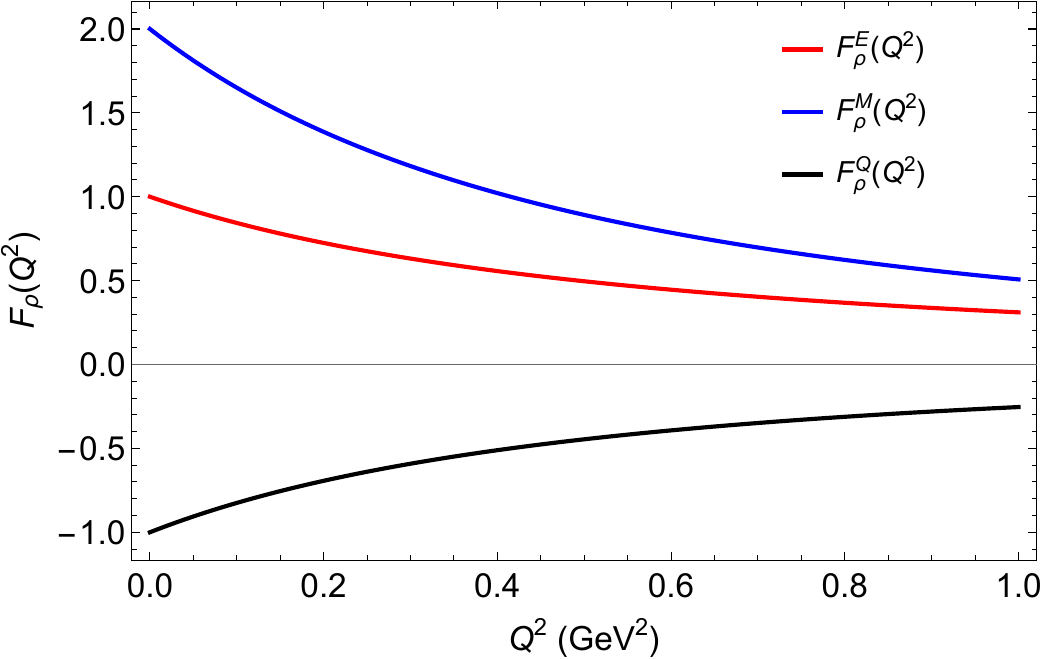}   
   \includegraphics[width=0.45\linewidth]{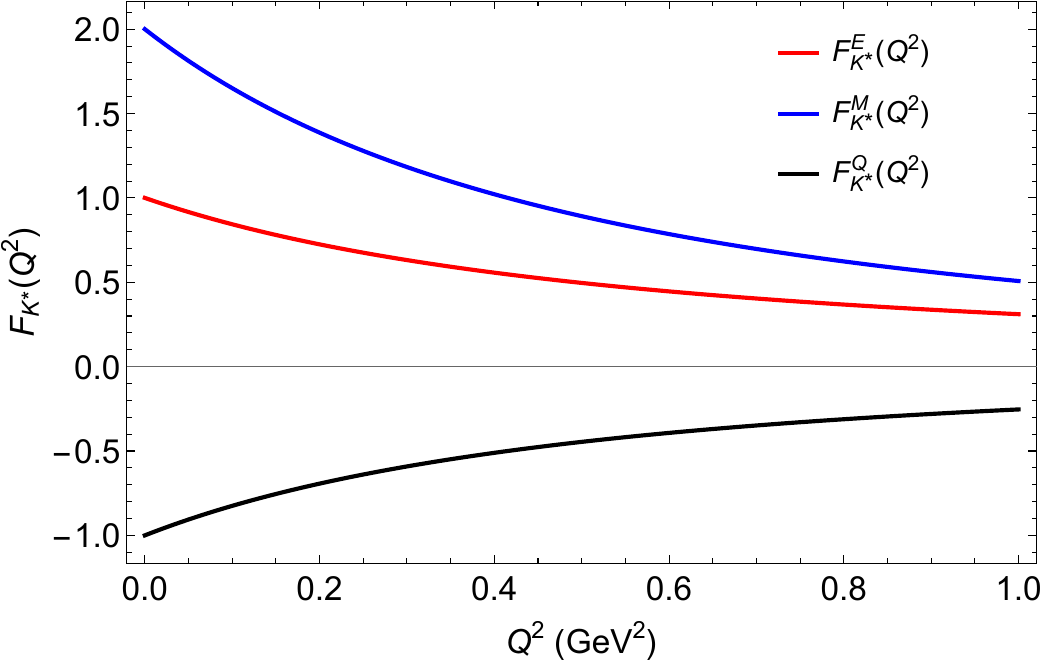} 
  \includegraphics[width=0.45\linewidth]{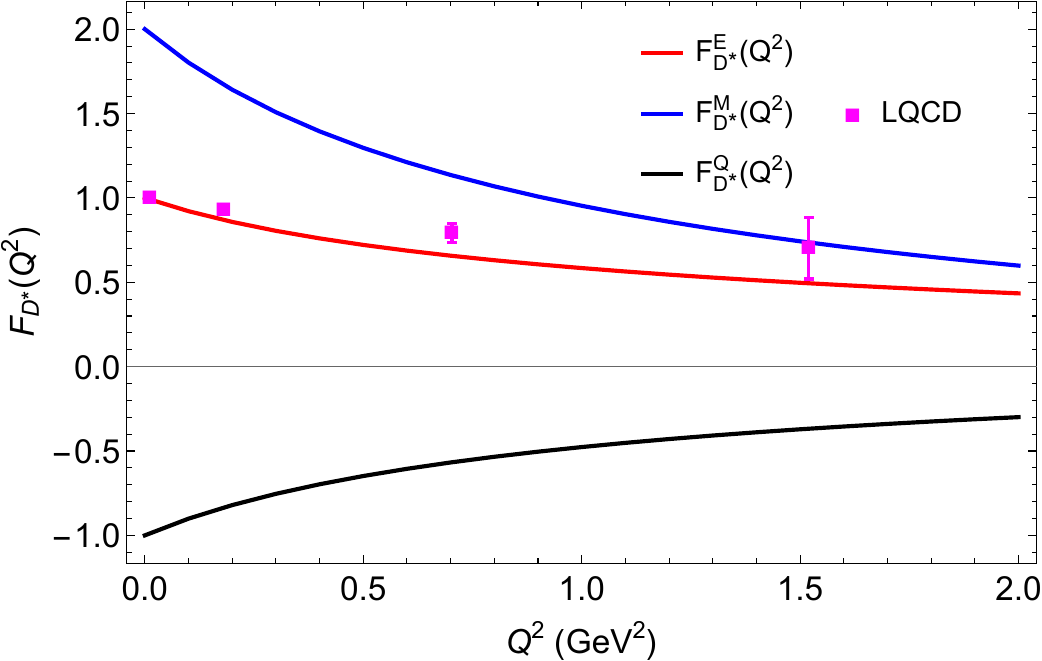}   
   \includegraphics[width=0.45\linewidth]{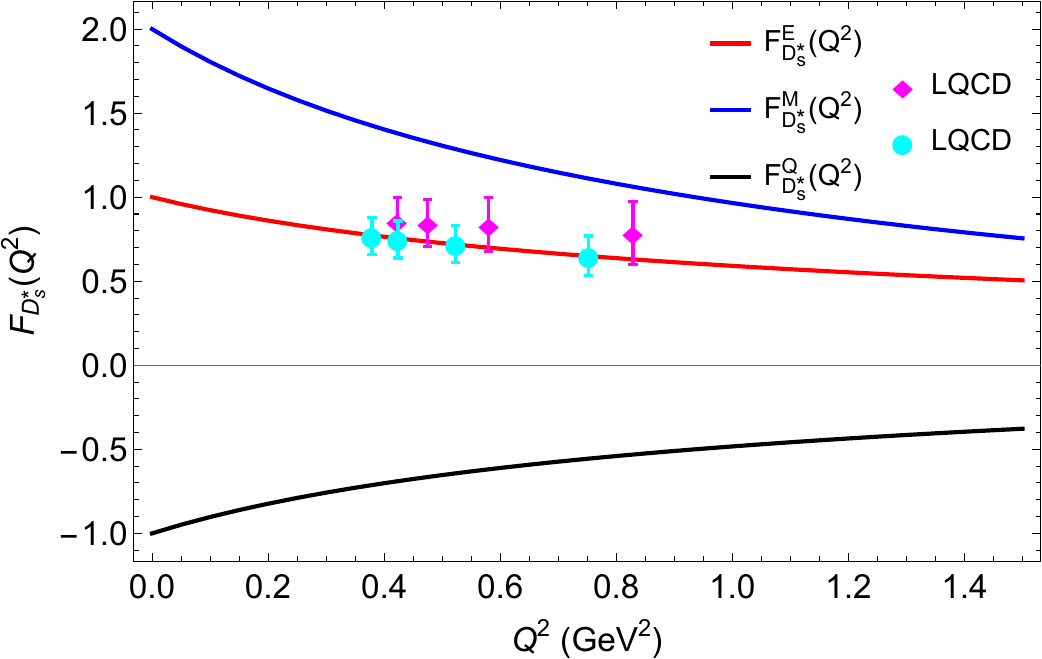} 
\caption{The electric(red), magnetic (blue), and quadrupole (black) form factors of the vector mesons ($\rho$: top-left, $K^{*}$: top-right, $D^{*}$: bottom-left, $D^{*}_{s}$: bottom- right). The lattice QCD data for the electric form factor of $D^{*}$, and $D_{s}^{*}$ are taken from Refs. \cite{Can:2012tx} with $m_{\pi}=300$ MeV and \cite{Cui:2019rid} for $m_{\pi}=300$ MeV (cyan points) and $m_{\pi}=315 $ MeV (magenta points), respectively.}
\label{formvector}
\end{figure} 
\begin{figure}
  \centering
  \includegraphics[width=0.45\linewidth]{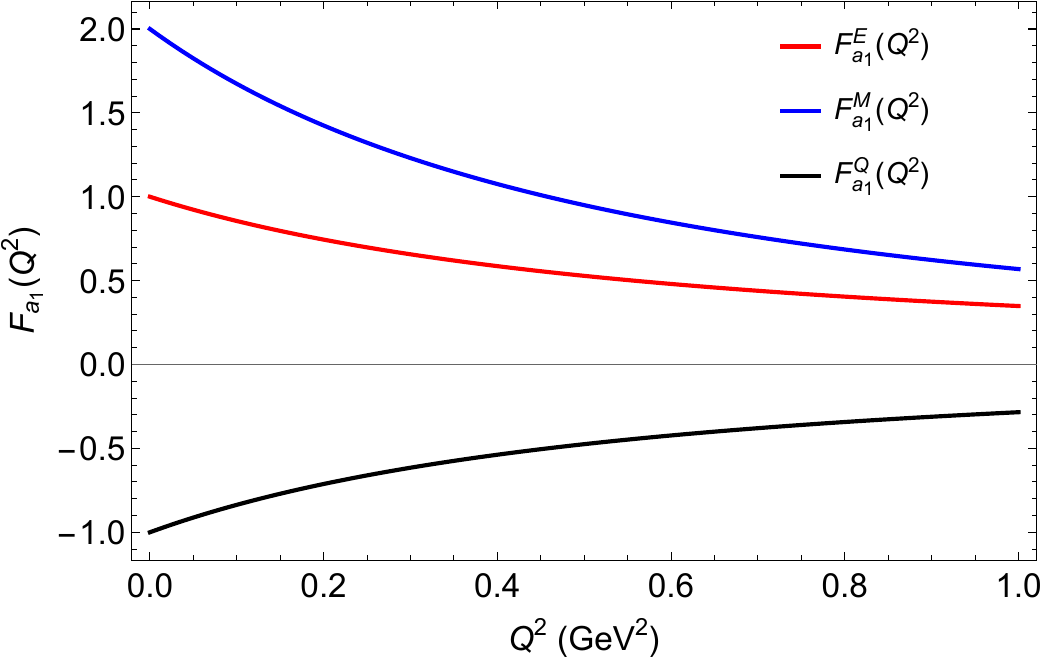}   
   \includegraphics[width=0.45\linewidth]{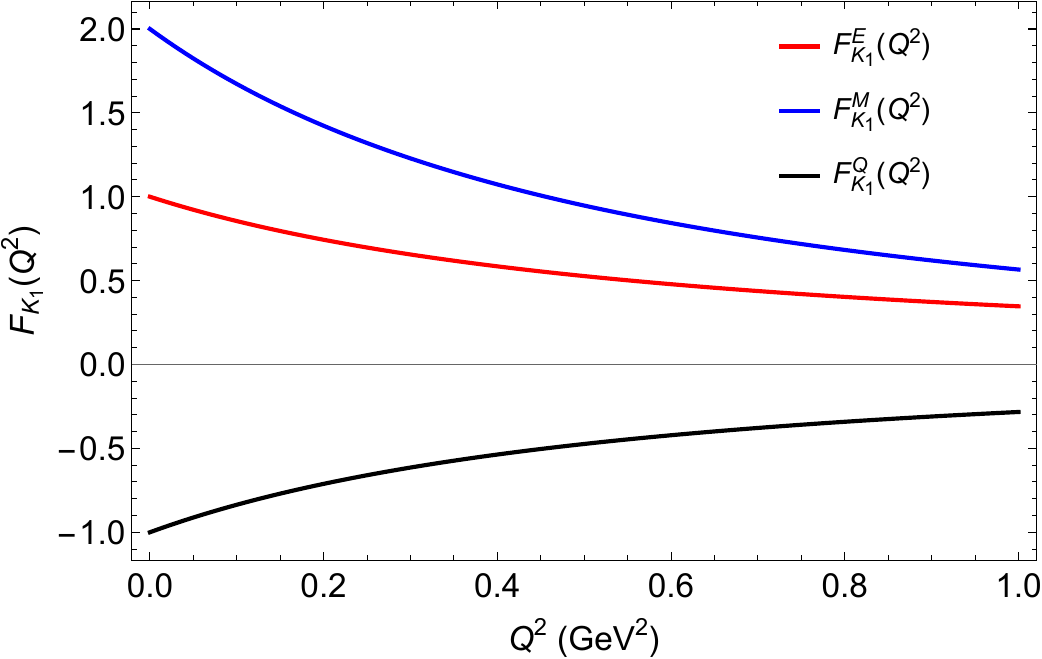} 
  \includegraphics[width=0.45\linewidth]{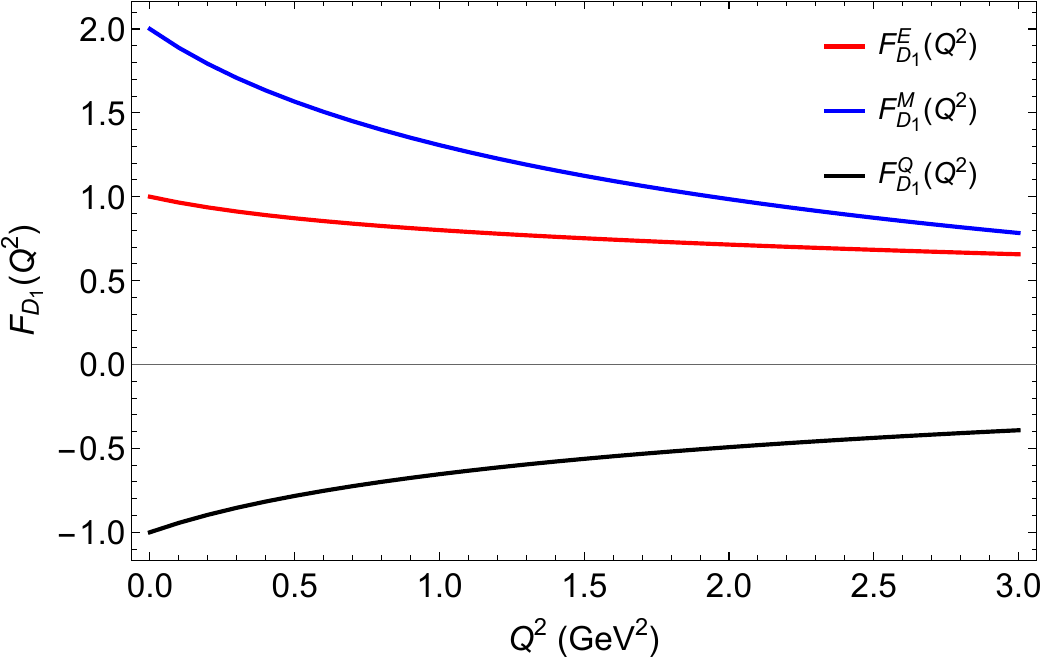}   
   \includegraphics[width=0.45\linewidth]{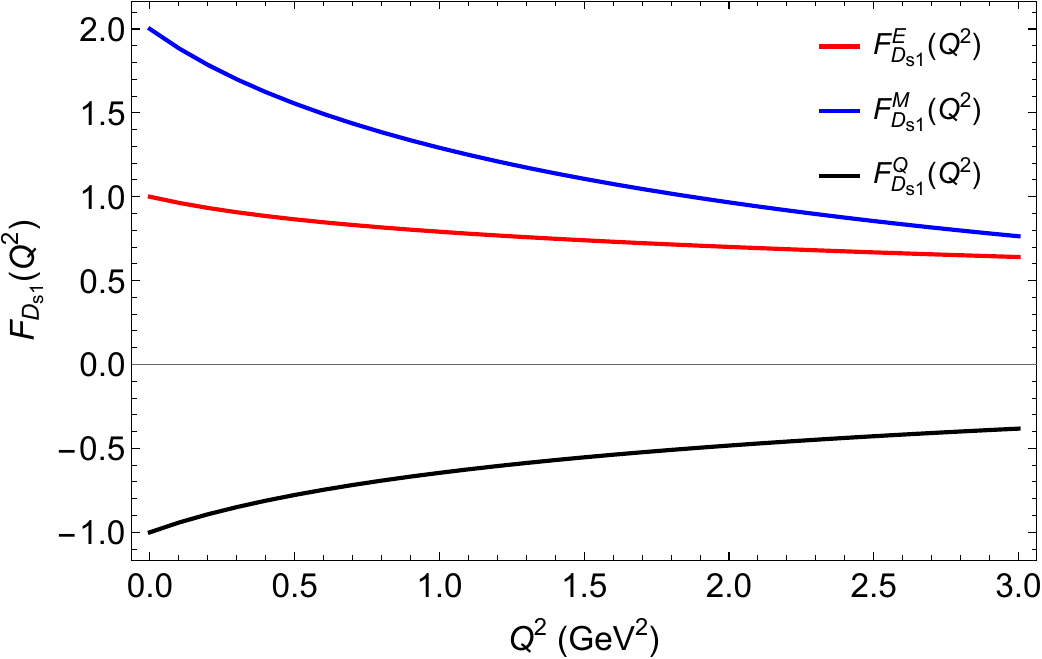} 
\caption{ The electric(red), magnetic (blue), and quadrupole (black) form factors of the axial vector mesons ( $a_{1}$: top-left, $K_{1}$: top-right, $D_{1}$: bottom-left, $D_{1s}$: bottom- right).}
\label{formaxial}
\end{figure} 
\begin{figure}
  \centering
  \includegraphics[width=0.45\linewidth]{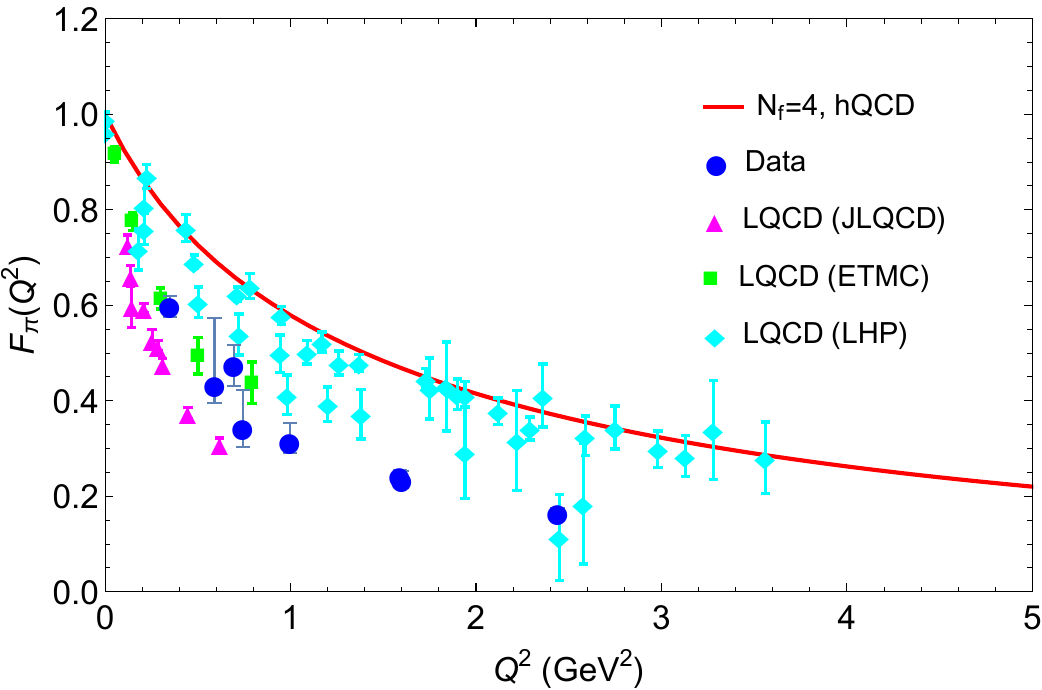}   
   \includegraphics[width=0.45\linewidth]{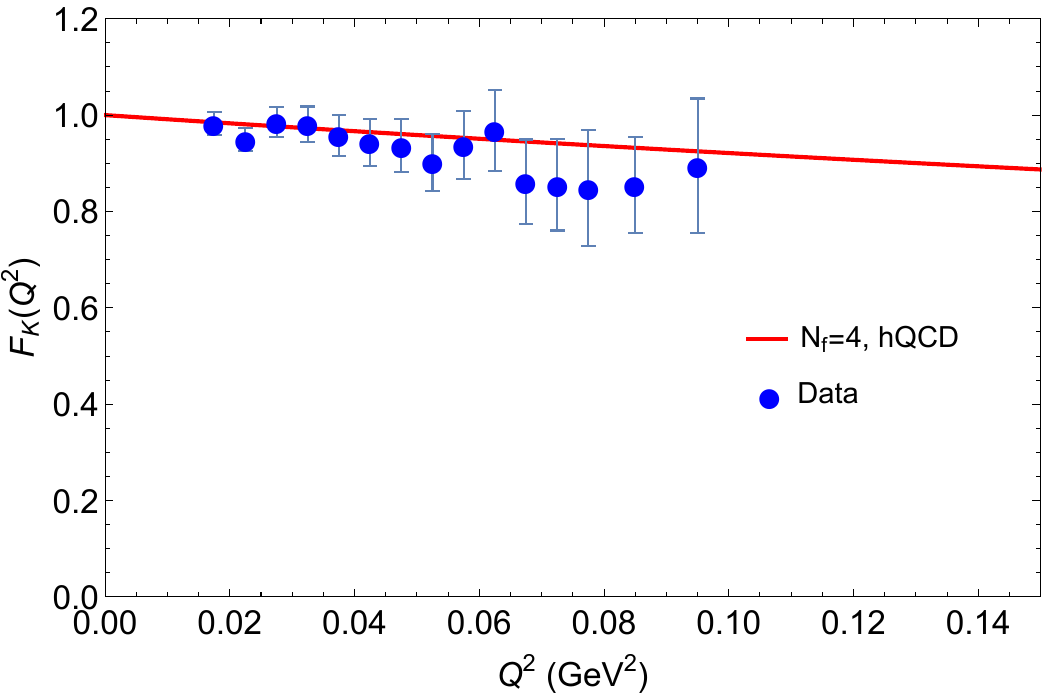} 
  \includegraphics[width=0.45\linewidth]{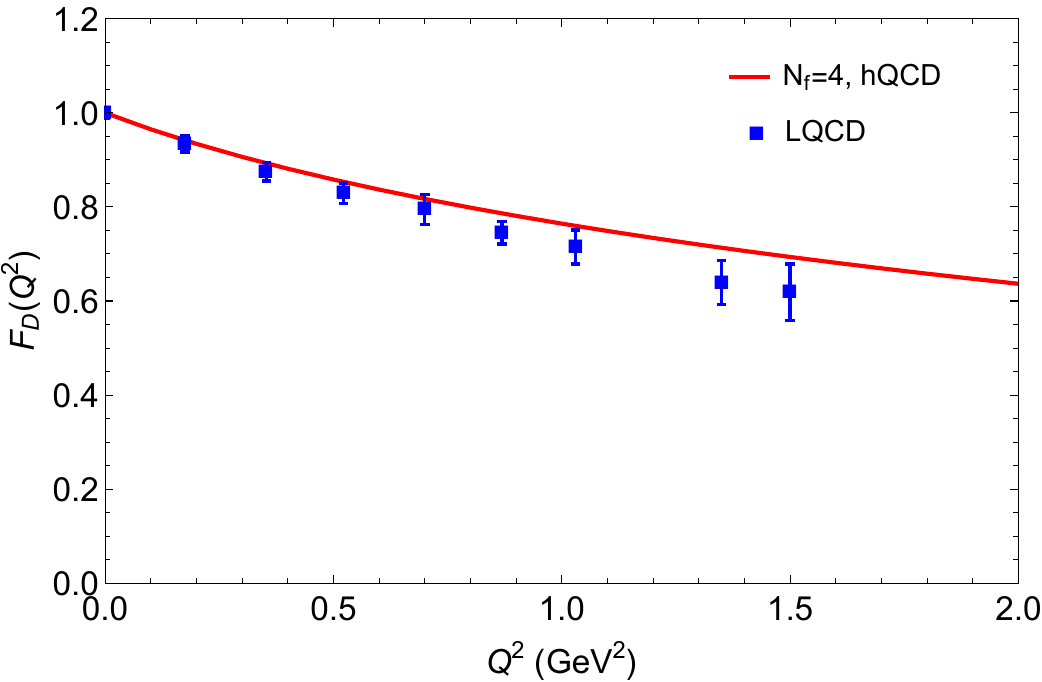}   
   \includegraphics[width=0.45\linewidth]{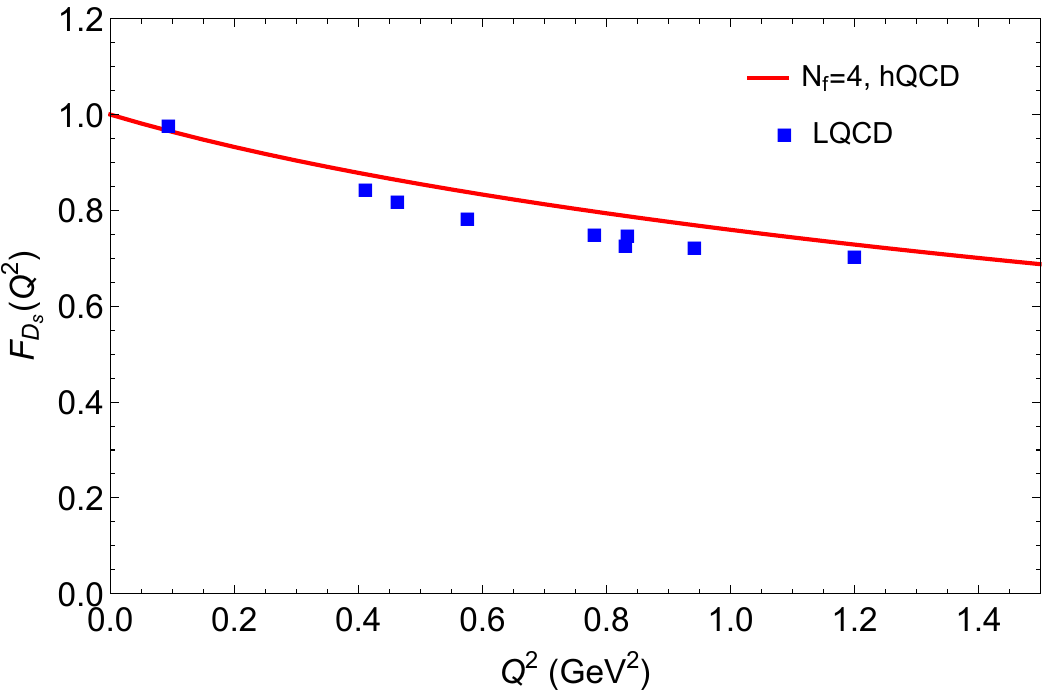} 
\caption{The form factors of the pseudoscalar mesons ($\pi$: top-left, $K$: top-right, $D$: bottom-left, $D_{s}$: bottom- right). The pion form factor is compared with the experimental data (Blue points)  \cite{JeffersonLab:2008jve} and LQCD data which are magenta points \cite{JLQCD:2008kdb}, green points \cite{Frezzotti:2008dr}, and cyan points \cite{Bonnet:2004fr}. The Kaon form factor is compared with the experimental data in Ref. \cite{Amendolia:1986ui}. The lattice results from which the $D$ and $D_{s}$ mesons are compared to are taken from Refs. \cite{Can:2012tx} and \cite{Li:2017eic}, respectively.}  
\label{formpion}
\end{figure} 

We present the results of the meson form factors obtained using the definitions of electromagnetic (EM) form factors described in the previous section. Firstly, we focus on the vector meson form factors and display them in Fig. \ref{formvector}. 
The form factors at $Q^2 = 0$ are well-defined for the ground states of the four vector mesons studied in this work, $\rho$, $K^{*}$, $D^{*}$, and $D_{s}^{*}$, such that $F_{V^{a}}^{E}(0)=1$, $F_{V^{a}}^{M}(0)=2$, and $F_{V^{a}}^{Q}(0)=-1$. Unfortunately, the experimental data is not available for the EM form factors of the vector mesons. However, one can find the lattice QCD result for the $D^{*}$ meson in the Ref. \cite{Can:2012tx} with $m_{\pi}=300$ MeV, and Ref. \cite{Cui:2019rid} provided two sets of solutions for the $D_{s}^{*}$ based on a different value of the lattice spacing and pion mass ($m_{\pi}=300$ MeV, and $m_{\pi}=315$ MeV). From Fig. \ref{formvector}, One can see that the electric form factors of the $D^{*}$ and $D_{s}^{*}$ mesons are compatible with the lattice QCD results. 

Similarly, we display the results of the electric, magnetic, and quadrupole form factors of the axial vector mesons in Figure \ref{formaxial}. Experimental data and lattice QCD results are currently unavailable for the axial vector sector. We hope that experimental and lattice QCD collaborations will report these results in the future. Our results for the axial vector mesons are comparable to those studied in the hard-wall holographic approach \cite{Ballon-Bayona:2017bwk}. Finally, we examine the form factors of the $\pi$, $K$, $D$, and $D_{s}$ mesons, and present the results in Figure \ref{formpion}. 

The pion form factor has been extensively studied from both theoretical and experimental perspectives. Previous works have investigated the pion form factor within the hard-wall and soft-wall holographic QCD models for the two-flavor case \cite{Gherghetta:2009ac,Grigoryan:2007wn,Kwee:2007nq,Chen:2022pgo}. Experimental data for the pion form factor have been reported by the Jefferson Lab collaboration for $Q^2 = 0.60 - 2.45$ GeV \cite{JeffersonLab:2008jve}. 

There are different lattice QCD results, we compare our results with the one reported by JLQCD \cite{JLQCD:2008kdb}, ETMC \cite{Frezzotti:2008dr}, and Lattice Hadron Physics (LHP) \cite{Bonnet:2004fr} collaborations. Figure \ref{formpion} demonstrates that our result is consistent with the lattice QCD result reported in Ref. \cite{Bonnet:2004fr}, although discrepancies with experimental data have been reported in earlier works within holographic QCD \cite{Gherghetta:2009ac,Grigoryan:2007wn,Kwee:2007nq,Chen:2022pgo}. Moreover, the comparison of the kaon form factor with the experimental data at low $Q^2$ \cite{Amendolia:1986ui} is shown in Fig. \ref{formpion} (top-left). One can see that our prediction is in very good agreement with the available data. For the case of the $D^{0}$ and $D_{s}$ mesons, only the data from the lattice QCD is available. The results of the EM form factors for $D^{0}$ and $D_{s}$ mesons are shown in Fig. \ref{formpion} (bottom-right) and (bottom-left), respectively. Our model's predictions are well consistent with the lattice QCD data.

From the results comparing with data, we can see that the form factors in the charm sector are in good agreement with experimental data, but for light mesons, especially for pion, the form factor deviates from the experimental data.

\subsection{Charge radii}
We calculate the charge radii of vector, axial vector, and pseudoscalar mesons in holographic QCD using Equation \eqref{chargeradii}. The results are presented in Table \ref{tab:chargeradius}. We compare the charge radii of the pion and kaon with the values listed in the Particle Data Group (PDG) \cite{ParticleDataGroup:2020ssz}. For the $D^+$ and $D^{*+}$ mesons, we compare our results with lattice data from Ref. \cite{Can:2012tx}. Similarly, we compare the $D_s^+$ and $D_s^{* +}$ mesons with lattice data from Refs. \cite{Li:2017eic} and \cite{Cui:2019rid}. However, there are no available experimental or lattice results for the remaining mesons' charge radii. It is seen that the charge radii of pion is in 65$\%$ agreement with the experimental data, and $D_s^{*+}$ meson is in 95$\%$ agreement with the experimental data.

\begin{table}
\center
\begin{tabular}{ c  c c  c  c }
\hline
\hline
          Observable                       &    & $N_{f}=4$, hQCD (fm) & &  Data (fm)    \\
\hline
          $r_{\pi^{+}}$       &    &  0.43     &  &    0.66    \cite{ParticleDataGroup:2020ssz}        \\ 
\hline
          $r_{K^{+}}$        &    &  0.45       &  &   0.56   \cite{ParticleDataGroup:2020ssz}         \\ 
\hline
          $r_{D^{+}}$        &    &  0.29      &  &    0.37 \cite{Can:2012tx}     \\ 
\hline      
          $r_{D^{+}_{s}}$    &    &  0.31      &  &    0.35 \cite{Li:2017eic}     \\ 
\hline      
          $r_{\rho^{+}}$     &    &  0.65      &  &              \\ 
\hline      
          $r_{K^{*+}}$       &    &  0.65      &  &               \\ 
\hline     
          $r_{D^{*+}}$       &    &  0.57      &  &    0.44 \cite{Can:2012tx}    \\ 
\hline      
          $r_{D_{s}^{*+}}$   &    &  0.46      &  &    0.44  \cite{Cui:2019rid}   \\ 
\hline      
          $r_{a_{1}^{+}}$    &    &  0.62      &  &               \\           
\hline      
         $r_{K_{1}^{+}}$     &    &  0.62      &  &               \\ 
\hline      
          $r_{D_{1}^{+}}$    &    &  0.28      &  &              \\ 
\hline      
          $r_{D_{s1}^{+}}$   &    &  0.32      &  &               \\ 
     \hline
     \hline
\end{tabular}
\caption{ Set of predictions for charge radius,
compared to experimental or lattice QCD data.}
\label{tab:chargeradius}
\end{table}

\section{Conclusions}
\label{conclusion}

In the present work, we performed the four flavors modified soft-wall holographic model to investigate various aspects of mesons, including meson spectra, decay constants, electromagnetic form factors, and charge radii of the vector, axial vector, and pseudoscalar mesons. The model parameters were fitted to experimental meson masses and decay constants, enabling us to obtain  the meson spectra for the vector mesons, $\rho$, $K^{*}$, $\omega$, $D^{*}$, $D_{s}^{*}$, and $J/\psi$, axial vector mesons, $a_{1}$, $K_{1}$, $f_{1}$, $D_{1}$, $D_{s1}$, and $\chi_{c1}$, and pseudoscalar mesons, $\pi$, $K$, $\eta$, $D$, $D_{s}$, and $\eta_{c}$. By considering the higher order terms in action in our model,  comparing the meson spectra with the results obtained in Ref. \cite{Chen:2021wzj}, our results are significantly improved, especially for the pseudoscalar sector. Moreover, the decay constants were also computed and reported in Table \ref{tab:decay}. In the vector sector, we predicted the decay constants of the $K^{*}$, $D^{*}$, and $D_{s}^{*}$ mesons and compared the result for the $\rho$ meson with experimental data, revealing a discrepancy of $16\%$. In the axial vector sector, the decay constant of the $a_{1}$ meson exhibited excellent agreement with the experimental data. Additionally, the decay constants of the pion and kaon were well reproduced in our model and compared with the experimental data, while the comparison for $D$ and $D_s$ mesons were made with lattice data. 

Considering the strong coupling constants of the $g_{\rho \pi \pi}$, $g_{\rho K K}$, $g_{\rho D D}$, $g_{\rho \rho \rho}$, $g_{\rho K^{*} K^{*}}$, $g_{\rho D^{*} D^{*}}$, $g_{\rho a_{1} a_{1}}$, $g_{\rho K_{1} K_{1}}$, and $g_{\rho D_{1} D_{1}}$, we investigated the $SU(4)$ flavour symmetry breaking. The results demonstrated that flavor symmetry is broken due to the different values of the quark masses and condensates, as indicated in Table \ref{tab:couplingratio}.

Furthermore, we studied the electromagnetic form factors and charge radii of the mesons of $\rho$, $K^{*}$, $D^{*}$, $D_{s}^{*}$, $a_{1}$, $K_{1}$, $D_{1}$, $D_{s1}$, $\pi$, $K$, $D$, and $D_{s}$. In the vector sector, our predicted electric form factors for $D^{*}$ and $D_{s}^{*}$ were found to be compatible with available lattice data. The predicted results of the axial vector mesons' electric, magnetic, and quadruple form factor are shown in Fig. \ref{formaxial}. For the pion meson, our work has a discrepancy with the experimental data; however, it is in agreement with the lattice results from Ref. \cite{Bonnet:2004fr}, as depicted in Fig. \ref{formpion}. Moreover, the kaon form factor exhibited good agreement with experimental data at low $Q^2$, similar to the results for $D$ and $D_s$ mesons, which also aligned well with lattice data. Finally, we calculated the charge radii using the holographic QCD. The obtained values are provided in Table \ref{tab:chargeradius}, where we predicted and compared the results with the available experimental or lattice data.

From these results, we can see that the physical quantities in charm sector is in good agreement with experimental data, but for light flavor mesons, the model predictions deviate away from the experimental data. This may indicate that the realization of chiral symmetry breaking in light flavor sector need to be improved further.

\section*{Acknowledgments}

This work is supported in part by the National Natural Science Foundation of China (NSFC) Grant Nos: 12235016, 12221005, 12147150, 12305136 and the Strategic Priority Research Program of Chinese Academy of Sciences under Grant No XDB34030000, the start-up funding from University of Chinese Academy of Sciences(UCAS), and the Fundamental Research Funds for the Central Universities. H. A. A. acknowledges the "Alliance of International Science Organization (ANSO) Scholarship For Young Talents" for providing financial support for the Ph.D. study.

 \addcontentsline{toc}{section}{References}
\end{document}